\documentclass{article}

\usepackage{microtype}
\usepackage{graphicx}
\usepackage{subfigure}
\usepackage{booktabs} 
\usepackage{amsmath}
\usepackage{tikz}
\usetikzlibrary{bayesnet}
\usetikzlibrary{positioning}

\usepackage{hyperref}

\usepackage[accepted]{icml2024}

\usepackage{amsmath}
\usepackage{amssymb}
\usepackage{mathtools}
\usepackage{amsthm}

\usepackage[capitalize,noabbrev]{cleveref}

\theoremstyle{plain}

\theoremstyle{definition}

\theoremstyle{remark}

\usepackage[textsize=tiny]{todonotes}

\icmltitlerunning{Position Paper: Why the Shooting in the Dark Method Dominates Recommender Systems Practice}

\begin{document}

\twocolumn[
\icmltitle{Position Paper: Why the Shooting in the Dark Method Dominates Recommender Systems Practice; A Call to Abandon Anti-Utopian Thinking}



\icmlsetsymbol{equal}{*}

\begin{icmlauthorlist}
\icmlauthor{David Rohde}{yyy}
\end{icmlauthorlist}

\icmlaffiliation{yyy}{Criteo AI Lap, Paris, France}

\icmlcorrespondingauthor{David Rohde}{d.rohde@criteo.com}

\icmlkeywords{Machine Learning, ICML}

\vskip 0.3in
]




\begin{abstract}
Applied recommender systems research is in a curious position.  While there is a very rigorous protocol for measuring performance by A/B testing, best practice for finding a `B' to test does not explicitly target performance but rather targets a proxy measure.  The success or failure of a given A/B test then depends entirely on if the proposed proxy is better correlated to performance than the previous proxy. No principle exists to identify if one proxy is better than another offline, leaving the practitioners \emph{shooting in the dark}.  The purpose of this position paper is to question this anti-Utopian thinking and argue that a non-standard use of the deep learning stacks actually has the potential to unlock reward optimizing recommendation.
\end{abstract}

\section{Introduction}
\label{intro}

The purpose of this paper is to address the curious position of applied, \emph{in production} recommender systems, and to acknowledge that practice is particularly poorly serviced by academic research.  Since at least the 2000s the industry has converged on a rigorous system to evaluate overall recommender system performance using online controlled experiments, usually refereed to as A/B tests \cite{kohavi2009controlled,kohavi2013online}.  A/B testing is rightly viewed as the `gold standard' for measuring actual performance, and its widespread adoption is correctly viewed as a great leap forward in using empiricism to measure  and optimize the performance of live systems (including recommender systems).  While a major breakthrough,  A/B testing based methodology leaves open: \emph{how to propose a new system to test?}

Unfortunately progress in building reward optimizing systems effectively stalled immediately after this innovation. The usual methods to propose a new recommender is to solve an optimization problem that targets only a loose proxy of performance (typically by proposing a user-item distance or even an item-item distance).  The notion of `distance' is quantified  using another dataset entirely, usually collaborative filtering or content, and heuristics are used to convert the `distance' into an implementable recommendation algorithm. As the success of such a methodology is completely dependent on the proposed proxy between this distance, and actual observed performance holding at A/B test time - we call this the \emph{shooting in the dark method}.  



A/B testing splits users into two (or more)  groups.  Users in group`A'  are then exposed to one intervention (or policy), and users in group `B' to another.  It is then reasonable to conclude that any difference observed  is due to the different recommendation policies; subject to just a few relatively benign assumptions.  While the inspiration of A/B testing is drawn from randomized control trials in medicine,  unlike randomized control trials, the policies `A' and `B' are potentially complicated interactive systems; where  in medicine `A' and `B' might be very simple, perhaps the administration of a drug or a placebo.  

Drawing valid conclusions from A/B testing relies on a number of assumptions. One important assumption is stationarity of both the type of users arriving, and the response of these users to actions.
A further and more subtle assumption is the Stable Unit Treatment Value Assumption (SUTVA) which simply refers to the idea that treatments applied to users in group `A' cannot have any impact on users in group `B'
\cite{rubin1986statistics,imbens2010rubin}.  Where in medicine SUTVA usually holds without exception: giving a drug to a person in group `A' has no impact on the people in group `B' who receive a placebo.  In recommender systems this assumption might be questionable, for example if a person in group `A' speaks to a person in group `B' perhaps influenced by a recommendations they received then this would violate SUTVA.  A more serious violation would occur when a single individual access the recommender system using different accounts and is therefore exposed to both system `A' and system `B'.  A well designed A/B testing protocol will seek to minimize SUTVA problems by either discouraging multiple accounts or tracking the user correctly over multiple devices.  In cases where this is not possible this will result in compromised causal inference, typically the outcome will be that the observed difference between `A' and `B' will underestimated as some users are exposed to both systems.


A first step in building reward optimizing machine learning systems is connecting the details of the user timelines with the highly aggregated and simplified A/B test results.  An appropriate framework is developed from first principles in Section \ref{timelinetoabtest}.  While, the principles of a full Bayesian decision theoretic approach is straightforward\cite{lad1996operational,bernardo2009bayesian,robert2007bayesian,de1937prevision,kadane2020principles}, implementation requires solving substantial
inferential problems and optimization problems.  In Section \ref{contextualbandit} it is proposed that while a contextual bandit framing is an imperfect model for full user trajectories, there are two principles which make it feasible to use on real world systems.  In Section \ref{contextbanditstillhard} it is explained why solving recommendation problems directly in the contextual bandit framework is difficult in practice, specifically covering high dimensionality, small effects, and causality considerations.  At this stage we are finally able to explain why the shooting in the dark method is so popular in Section \ref{shootinginthedark}.  Then
Section \ref{pclick}  discusses the widespread use of large scale (maximum likelihood) logistic regression (and other generalized linear models), while this approach is powerful it is again argued that
 this approach too cannot be considered a direct attack on reward optimizing recommendation.    In Section \ref{deepbutnottogodeep} a concrete proposal is made for a direct attack on reward optimizing recommendation via an extended contextual bandit formulation, based on deep learning tooling (but not deep models).  Finally Section \ref{conclusion} concludes.

\section{From User Timelines to A/B Test Results}
\label{timelinetoabtest}

\subsection{Bayesian Decision Theoretic Recommendation}
A direct Bayesian decision theoretic formulation of reward optimizing recommendation is surprisingly straightforward, at least \emph{in principle}.  This will be done from first principles and results in a formulation somewhat more general than the Markov decision process usually adopted in reinforcement learning.

Initially  user u is observed in an initial state $S_0^u$, The initial state will have a marginal distribution $P(S_0=s_0^u|\theta_0)$.  
The information within $S_0$ contains all the information up until the first \emph{decision point}.  At this time a decision must be made by selecting the value of the first action $A_1$, this can be influenced by the information contained in the initial state $S_0$, if this decision rule is deterministic the decision rule can be written as $A_1=f_1(s_0)$.  The second state $S_1$ is then conditional on both the action and the previous state, i.e. $P(S_1=s_1^u|S_0=s_0^u,A_1=a_1^u,\theta_1)$.

The same principle continues until the final step ($T$)  i.e. $P(S_T=s_T^u|S_{0:T-1}=s_{0:T-1}^u,A_T=a_T^u,A_{0:T-1}=a_{0:T-1}^u,\theta_T)$, where $T$ itself may be a random variable, (and may change dependent on the user, in which case it is denoted $T_u$).  In developing the theory for reward optimizing recommendation, it is useful to be able to alternate between the actions actually delivered $A_{1:T}=a_{1:T}$ and the policy that delivers those actions $f_1(\cdot), ..., f_T(\cdot)$, which for brevity will written $f_{1:T}(\cdot)$.  Using this notation  the model for the joint of the outcome $s_{0:T}$ conditional on the policy $f_{1:T}(\cdot)$ is:

\begin{align}
\label{model}
P&(S_{0:T}=s_{0:T}^u|A_{1:T}=f_{1:T}(\cdot)) = \\ \nonumber
& P(S_T=s_T|A_{1:T-1}=f_{1:T-1}(s_{0:T-1}),S_{0:T-1}=s^u_{0:T-1},\theta_T) \\ \nonumber
& \times..\times P(S_2=s_2^u|A_{1:2}=f_{1:2}(s_0,s_1),S_{0:1}=s_{0:1},\theta_2) \\ \nonumber
& P(S_1=s_1^u|A_1=f_1(s_0),\theta_1,S_0=s_0^u)P(S_0=s_0^u|\theta_0) \nonumber
\end{align}

The posterior follows immediately:

\begin{align}
\label{posterior}
P&(\theta|\mathcal{D}) \propto  P(\theta_0,...,\theta_T) \prod_u \\
\nonumber
& \times P(S_T=s_T|A_{1:T-1}=a_{1:T-1},S_{0:T-1}=s^u_{0:T-1},\theta_T) \\ \nonumber
& \times...\\  \nonumber
& \times P(S_2=s_2^u|A_{1:2}=a_{1:2},S_{0:1}=s_{0:1},\theta_2) \\ \nonumber
& \times P(S_1=s_1^u|A_1=a_1,\theta_1,S_0=s_0^u,\theta_1) \\ \nonumber
& \times P(S_0=s_0^u|\theta_0) \nonumber
\end{align}

There are a few useful remarks to make about these equations.  Firstly, the joint on $S_{0:T}$ in Equation \ref{model} is conditional on the policy of how actions are delivered, not the specific actions themselves as in Equation \ref{posterior}.  This is necessary because the actions are determined by the behavior of earlier states.  In contrast, the likelihood and consequently the posterior (Equation \ref{posterior} ) uses the delivered actions, indeed if $f_1^A(s_0)=f_1^B(s_0)$, for some or all values of $s_0$, then data gathered from $f_{1:T}^A(\cdot)$ will provide (some) information about how $f_{1:T}^B(\cdot)$  behaves.

By conditioning the posterior on past data $\mathcal{D}=\{ s_{0:T_u}^u,a_{1:T_u}^u\}_{u=1}^M$    the predictive distribution can be computed: $P (S_{0:T}=s_{0:T}|f_{1:T}(\cdot),\mathcal{D})$.  Leaving aside the inferential difficulties hinted at earlier, we  proceed to a decision analysis by further specifying a utility function $U(s_{0:T})$.  The optimization problem  is then simply: \footnote{This useful formulation will be the focus of this discussion, but it is worth noting that in some cases it is necessary to not only define utilities over a single user as above, but over many users, in which case the utility must be defined $U(s_{0:T}^1,...,s_{0:T}^N)$.  A typical use case is when targeting a user depletes a budget and so the utility is given by the total value for a fixed budget.}:

\begin{align}
    \label{opt}
    f&_{1:T}^*(\cdot) = {\rm argmax}_{f_{1:T}(\cdot)} \\ \nonumber
    & \int U(s_{0:T}) P (S_{0:T}=s_{0:T}|A_{1:T}=f_{1:T}(\cdot),\mathcal{D}) d s_0 ... d s_T  
\end{align}

Despite this extremely simple formulation, implementing both the inference step in Equation \ref{posterior}, and the decision making step in Equation \ref{opt} is difficult in practice, and to the best of our knowledge, never seriously attempted.  A  core difficulty is that the likelihood in Equation \ref{posterior} is flat over most of the parameter space.  This is driven by two considerations, firstly if a deterministic policy  $f_{1:T}(\cdot)$ is used to collect the data, then there will be a fixed correspondence between $A_1=f_1(\cdot),...,A_T=f_T(\cdot)$, which will result in a lack of identifiablity of the parameters.  A randomized logging policy mitigates this, but only slightly.  
Alternatively, if an A/B test is being run of two competing policies $f^A_{1:T}(\cdot)$, or $f^B_{1:T}(\cdot)$, then this enables exploration of two quite distinct policies.  A further problem is simply the extreme dimensionality of the parameters $\theta_1,...,\theta_T$.  Inference deeper into the sequence becomes increasingly difficult  as the cardinality of the conditioning set multiplies with each likelihood term.  Even in the case of completely random (uniform) exploration the likelihood on $\theta_T$ will fail to concentrate in real world scenarios.  From a Bayesian perspective, \emph{the only way out of these difficulties} is to augment the admittedly very modest likelihood information with prior information.  

Turning our attention now to optimizing the decision problem in Equation \ref{opt}.  This problem is also complicated to solve as it incorporates forward planning considerations and solving it requires advanced techniques from either dynamic programming or reinforcement learning e.g. \cite{bellman1966dynamic,williams1992simple,sutton2018reinforcement}.  Although, the literature contains relatively little work on this problem due to the formidable nature of the inference problem.

\subsection{From the Model to the A/B Test Result}

A difficulty with the usual presentation of A/B testing is that it lacks a connection to the underlying model containing the details of the user trajectories, i.e. Equation \ref{model}.  All the details that might assist in evaluating an arbitrary policy are dropped from the formulation.  An arbitrary policy contains all the granular information about if a user ends up in a particular state, what action (recommendation(s)) should be delivered to them for \emph{every} possible state.  In contrast A/B testing is based on highly aggregate statistics.  This leaves us with no means to directly optimize $f_{1:T}(\cdot)$, or indeed propose a new $f^C_{1:T}(\cdot)$ for measurement.

A model conditional on a policy produces a full joint distribution on all the states i.e.  $P(S_{0:T}=s_{0:T}^u|f_{1:T}(\cdot))$, and implicitly the actions as well.  A/B testing should be viewed as willfully committing the Bayesian sin of \emph{marginalizing before conditioning}.  That is instead of considering the full trajectory of the states and the actions of users, all these details are neglected and the model only predicts the observed utility conditioned on the policy i.e. $P(U(s_{0:T})|f^A_{1:T}(\cdot))$, and $P(U(s_{0:T})|f^B_{1:T}(\cdot))$.  Once this is done it is possible to estimate the utility under each policy and do a very simple error analysis.  Of course this type of analysis throws away all the information in the user trajectories.

\section{Contextual Bandits}
\label{contextualbandit}
A Contextual Bandit is the problem described in Equation \ref{model} in the special case where $T=1$.  This case has both pleasant mathematical properties and limitations.


\subsection{Mathematical Properties}
Under a contextual bandit assumption the model becomes:

\begin{align*}
P&(S_1=s_1^u,S_0=s_0^u|f_1(\cdot),\theta_1,\theta_0)  \\
&= P(S_1=s_1^u|S_0=s_0^u,A_1=f_1(s_0^u),\theta_1)P(S_0=s_0^u|\theta_0)    
\end{align*}

Although the restriction to $T=1$ obviously limits the scope of a contextual bandit, some very useful mathematical results flow from this restriction.  Firstly the most important part of the model: $P(S_1=s_1^u|S_0=s_0^u,A_1=a_1)$ is just a, non-standard, regression or supervised learning problem.  Non-standard because $s_1$ might be a complex object.  It can be made a completely standard by regressing directly on the utility $v$:

\[
E[v|S_0=s_o,A_1=a_o,\theta_1].
\]
Although, the possibly useful information beyond the utility within $s_1$ is lost (again, conditionality is violated).  Moreover, the determining of the optimal action is simply:

\[
f_1^*(s_1) = {\rm argmax}_a  E(v|S_0=s_o,A_1=a_o,\theta_1) 
\label{banditopt}
\]
Moreover, depending on the paramterization it may be immediately construct an optimal $f_1(\cdot)$ for all values of $S_1=s_1$ from an estimate of $\theta_1$.

\subsection{Relevance of Contextual Bandits to Real Systems}

While restricting to a contextual bandit has pleasant mathematical properties, it is also intrinsically limited.  A real systems does not conform to the $T=1$ formulation.  Two methods that can be used to coerce the contextual bandit formulation to apply to real systems ($T>1$), the Meta-Policy approach and attribution approach.

\subsubsection{Make the decision rule a policy (Meta-Policy)}
Where the A/B testing approach takes  $P(S_{0:T}=s_{0:T}^u|f_{1:T}(\cdot))$ and marginalizes down to utility, producing a non-contextual bandit, allowing the estimation of policies.  The Meta-Policy turns the non-contextual bandit into a contextual bandit by conditioning on $S_0=s_0$ $P(S_{1:T}=s_{1:T}^u|f_{1:T}(\cdot),S_0=s_0)$, and treats the policy as the action \cite{betlei2023maximizing}.  Then the same pre-data marginalization compromises used in A/B testing are made i.e. the further details of the user trajectory from $s_{1:T}$ are ignored.  The practical impact of this is that after an A/B test testing two policies $f^A_{1:T}(\cdot)$, and $f^B_{1:T}(\cdot)$, the initial state $s_0$ can be used in order to determine pockets to use each of these policies.

The main benefit of this method is that it allows a (limited) level of personalization to follow after an A/B test, and, like A/B testing makes only very limited assumptions.  While there is a technical issue as there is a much higher error when comparing the non-personalized  $E(v|f^A_{1:T}(\cdot))-E(v|f^B_{1:T}(\cdot))$ with the personalized $E(v|f^A_{1:T}(\cdot),S_0=s_0)-E(v|f^B_{1:T}(\cdot),S_0=s_0)$, that issue aside this methodology can be as robust as A/B testing.

In contrast there are some important limitations: the action space is very large and exploring all but a few policies is infeasible, and personalization is only on the initial state $s_0$, 


\subsubsection{Attribute Value To Decision Points}

Unlike the Meta-Policy approach, attributing value back to decision points requires stronger modelling assumptions and a restricted form of the utility function.  Specifically let the utility function have the form:
\begin{align}
U(s_0,...,s_T) = \sum_i v(s_i),
\label{attributility}    
\end{align}

and, the model :
\begin{align}
\label{modelattrib}
P&(S_{r}=s_{r}|A_{1:r}=a_{1:r}^u,S_{0:r}^u=s_{0:r-1}^u) \\
& = P(S_{r}=s_{r}^u|A_{r}=a_{r}^u,S_{0}=s_{0}^u), ~ \forall_r. \nonumber
\end{align}

These assumptions while useful, are restrictive.  The utility function in Equation \ref{attributility} might be appropriate where the utility measures the number of items clicked, it can be interpreted as having one user click twice is equivalent to having two users click once each.  Attribution is good at handling click type metrics because the reward and the recommendation have a tight coupling, it handles much more poorly long term measures of long term user satisfaction such as sales, watch time.  This is because these rewards do not in a meaningful way `belong' to a recommendation decision point.

The modelling assumptions within Equation \ref{modelattrib}  corresponds to the situation where interests are contained in $s_0$, and $s_{1:T}$ contain only the response to the recommendation $a_{1:T}$, but no further information about the users interests.  

This formulation makes non-trivial assumptions, but the power of these assumptions is that every decision point within every user timeline can be viewed as a realization of a contextual bandit, both simplifying the model and increasing the data available for estimation (there are now $T_u$ records for each user $u$).

\subsection{Digression on attribution}

Attribution is a radical simplification of the full sequential inference problem in
 Equations  \ref{posterior} and \ref{opt}, yet we will see even this simplified problem is difficult to solve.  Coercing a complex recommender system trajectory into a contextual problem while imperfect is a promising way to make progress in the area of reward optimizing recommendation.  The fundamental problem is that the reward is associated with the user, but attribution requires it being `attributed' back to decision points. This may give the impression that we are in agreement with the Marketing Science Institute which has repeatedly described attribution as a top or even the top research priority in marketing\cite{msi}, or even attempts to axiomize  attribution \cite{singal2019shapley}.  Rather, to paraphrase Wolfgang Pauli we argue that
, attribution  \emph{is not even wrong} \cite{peierls1960wolfgang}.

It isn't too unreasonable to attribute a `click' to a (slate of) recommendation(s) that were or were not clicked.  This isn't entirely perfect.  Perhaps unclicked ads early in the sequence `nudged' the user to eventually click at a later stage in the sequence, or perhaps the opposite a user will become fatigued.  In any case there is no correct way to deal with these affects by `smart attribution'.  These problems become far more acute in the common case when the reward is not clicks, but something without an obvious connection to the recommendation, think, sales, views, watch time etc, and in most real systems these metrics are the true measures of performance.

Unfortunately there is no `correct' way to coerce the sequential learning and optimization problems in Equation \ref{posterior} and \ref{opt} to a contextual bandit using a smart axiomization of attribution.  It isn't possible to formalize attribution with a \emph{what if} formulation, to do so would suggest that we could view many realizations of a particular sequence of actions, and a counterfactual sequence perhaps with a single modification.  It is telling that in medicine it is considered incorrect to attribute an adverse effect to a definite cause, but this sort of reasoning is common in digital marketing.


\subsection{Reinforcement Learning and Other Timeline Aware Approaches}

Some researchers and practitioners have developed reinforcement learning approaches for recommendation.  Where Bayesian decision theory is founded in an axiomization of decision making under uncertainty, the field of Reinforcement learning (RL) \cite{sutton2018reinforcement} is not so easy to classify.  Model-free reinforcement learning can loosely be thought of as decision rule optimization, i.e. algorithms to solve Equation \ref{opt}  \cite{williams1992simple,watkins1992q,sutton2018reinforcement}.  RL approaches are particularly well suited to game playing where it is sufficient to have only noisy estimates of the utility to optimize solve Equation \ref{opt} and it doesn't matter if poor decisions are made in early stages of the optimization process.

While there are numerous proposals to apply reinforcement learning  to recommendation and hence go beyond the myopic contextual bandit \cite{chen2019top,ie2019slateq,chen2021survey}.  These methods invariably skip the subtle modelling and inference step in Equation \ref{posterior}, and either propose a) running directly on live traffic; or b) using estimators that perform catastrophically in far simpler situations.

Direct optimization on live traffic is an idea with some merit, but with the obvious downside of allowing the algorithm to perform poorly on live traffic. Fundamentally such an approach is a variant of A/B testing.  In contrast, offline RL approaches that shun the modeling problem in Equation \ref{posterior}  by using crude off policy estimators should be treated with  suspicion.  Achieving sufficiently accurate estimation in order to distinguish different sequential strategies seems ludicrously ambitious even when all the available information is used.


\section{Regression for Solving the Contextual Bandit}
\label{contextbanditstillhard}

\subsection{Estimation is (Unfeasibly) Hard}

At this stage we have made the case that reward optimizing recommendation can be achieved (imperfectly) by formulating the problem as a contextual bandit formulation (say) using the attribution heuristic.  This turns the learning problem into a regression or supervised learning task, and  decision making similarly can be achieved simply by an argmax over the recommendation space.  It would seem we have successfully reduced the problem down to one that can be attacked with well understood tooling such as large scale logistic regression or deep supervised learning.  Unfortunately, \emph{this is false}.  This is false, even in toy recommendation situations where the set of recommendable items is a single item (rather than the slate or banner typical in real environments).

In order to demonstrate this, let's take a truly toy recommender system. Imagine that $S_0$ is a discrete random variable with 1000 topics that the user might be interested in.  Similarly imagine that the recommendation set or actions are also drawn from a discrete set with 1000 possible values, and $S_1$ is a Bernoulli variable that identifies if the item was clicked or not.  Furthermore imagine that the click through rate of the best action is $0.9\%$ which is the case for exactly one action for each context, and the other actions have a click through rate of $0.1\%$.  In order for reduce the standard error of the difference between the best action and another action down to $\pm 0.1\%$, this requires $10 0000$ observations on each context-action pair.  There are also $10 0000$ context action pairs.  If the logging policy or past recommender system is purely random then to achieve this would require $10^8$ records i.e. 100 million records.  The assumption that the logging policy is uniform is very unrealistic, let's assume instead that there is an epsilon greedy policy running, this will choose whatever the previous system thinks is best $99\%$ of the time and explores $1\%$ of the time.  This further increases the number of records required to a very sizeable 10 billion. In other words, absurdly large datasets are required even for toy problems.  In real systems the action size is likely to be overwhelmingly larger than 10000, not (only) because of large real world catalogue sizes, but because many items are shown at once in a banner (or slate) resulting in a combinatorial explosion in the action size which will multiply the 10 billion by something very large indeed. 


Our reduction of the recommendation modelling problem via the contextual bandit down to a simple regression problem is unfortunately a Pyrrhic victory.  The dimension of the interaction effects between the context and action is enormous even in completely toy problems, and the effect size so small that the amount of data required in order to produce reasonable estimates for all context-action pairs is completely out of reach.
The consequence of all of this is simply to acknowledge that accurate estimation for all context-action pairs in recommendation settings based on the reward signal alone is completely out of reach.  \emph{While the reward signal is of paramount value, using it on its own for every context-action combination is a fool's errand.}

\subsection{Causation is Trivial}

After this, sobering perspective on the difficulty of estimation we present some good news. \emph{Causal inference is trivial}.  Indeed, in general from a Bayesian decision theoretic position, causal inference is inference and there is no need to augment it with exotic extensions \cite{rohdecaus,rohde2022causal,lattimore2019replacing} such as the do-calculus \cite{pearl1995causal} or the Rubin causal model\cite{rubin1986statistics}, or non-stochastic regime indicators \cite{dawid2021decision}.  Moreover, when the model is a contextual bandit simple regression is sufficient to  immediately obtain correct causal inference.

\subsubsection{Unobserved confounding won't occur (unless we confound ourselves)}

Several papers claim that in order to build effective recommender systems we need to incorporate \emph{causal reasoning}.  One example is::

\begin{quote}
To estimate such a causal effect, it is thus essential to incorporate conventional causal inference techniques into recommender models. \cite{gao2022causal}   
\end{quote}


This seems to suggest that recommender systems practitioners have yet another technical field to master \emph{causal inference}.  This is untrue, and confusing.  Correct causal inference can be achieved simply with regression or supervised learning i.e. by so called `back-door adjustment' \cite{pearl1995causal}.  Unobserved confounding is a known concern in drawing causal conclusions from non-experimental data, but unobserved confounding occurs only when an un-observed variable determines both the allocation of the treatment, and the response to the treatment.  This cannot occur in a production system, because a production system exclusively uses observed variables to allocate treatments.  

On the other hand, it is possible for poor practices in the  real world systems to cause unobserved confounding.  In fact these situations and practices are common.  One situation occurs when features are dropped from a model.  Then  model to be trained takes the form: $P(S_1=s_1|S_0=s_0,a_1)$, but the \emph{past logging policy} is of the form $\pi(A_1=a_1|S'_0=s'_0)$, where $s'_0$ contains strictly more information than $s_0$.  This means that the action delivered, can provide information about the missing information in $s_0$ that helps predict the outcome $s_1$, but is not causal.  The usual approach to handling this sort of problem is to simply allow the confounding to temporarily occur.  As models are typically trained only on recent history after a short fixed delay the confounding will disappear again.

Another cause of unobserved confounding is due to nesting multiple models each using different information.  Consider the situation where $s_1$ contains both information about a click $c$ and a buy $b$.  Then this can be modelled:
\[
P(B=b,C=c|S_0=s_0,a_1)
\]
Actions, might be optimized in order to maximize some utility function e.g. post click sales, $U(B=1,C=1)=1$, and  $U(B=1,C=0)=U(B=0,C=1)=U(B=0,C=0)=0$.  This can be achieved by three logistic regression models: $P(C=1|S_0=s_0,a_1)$, and $P(S=1|S_0=s_0,a_1,C=1)$, $P(S=1|S_0=s_0,a_1,C=0)$. Although the third model isn't needed for this particular utility function.  Actions will then be determined by $a_1^* = {\rm argmax} ~ P(B=1,C=1|S_0=s_0,a_1)$.

So far so good.  However, in a large company the buy model and the click model are likely optimized by separate teams.  Each of these teams will have some very smart people who do clever feature engineering.  This results in two models $P(C=1|S_0'=s_0',a_1)$, and $P(S=1|S_0''=s_0'',a_1,C=c)$, where $s_0'$, and $s_0''$ are features tuned to optimize the performance of the click and sales model respectively.  The actions are now determined by:
$a_1^* = {\rm argmax} P(B=1|S_0''=s_0'',a_1,C=1)P(C=1|S_0'=s_0',a_1)$.  That is, the action is influenced by both $s_0'$, and $s_0''$, and therefore contains information about both causing both the buy and click model to suffer from unobserved confounding.


The remedy to the unobserved confounding problem isn't to understand some abstract theory of causality such as structural causal models \cite{pearl1995causal}, but a rather simple one.  All models, at every level of the hierarchy should have access to exactly the same information, or there should only be one end-to-end model.

\subsubsection{Off Policy Estimation with the Horvitz-Thompson Estimator}

Another method is often used under the name `causal inference' for contextual bandits, but in truth has more to do with computing an average treatment effect.  Within the contextual bandit framework there are two important calculations that a decision maker or recommender can consider.  The first is given an initial context $s_0$, what is the optimal decision, this results in a perfect policy/recommender (assuming the model is correct) by implementing Equation \ref{banditopt}.

Instead of going directly to the optimal policy, it is possible to simply evaluate the utility of an arbitrary policy $f(\cdot)$.  One reason for considering non-optimal polices is that not all policies are implementable in practice; e.g. solving the argmax within Equation \ref{banditopt} might be infeasible.  The expected utility of an arbitrary policy is often called the value function and denoted $\mathcal{V}(\cdot)$, and can be computed using:

\begin{align}
    \label{valuefunc}
\mathcal{V}&( f(\cdot) ) = \\ \nonumber
& \int U(s_1) P(S_1=s_1|A_1=f(s_0),S_0=s_0) \\ \nonumber
& ~ \times P(S_0=s_0) d s_0 d s_1
\end{align}

It then becomes possible to optimize $f(\cdot)$ with respect to $\mathcal{V}$.  If the policy space is unconstrained then the  marginal distribution $P(S_0=s_0)$ is not needed, as the optimal action is taken for every $s_0$.  When the policy space $f(\cdot)$ is restricted then $P(S_0=s_0)$  becomes important, as it is better to allocate sub-optimal actions to rare values of $s_0$ if it allows delivering near-optimal ones on common values of $s_0$.

Computing $\mathcal{V}(\cdot)$ is straightforward from a Bayesian decision theoretic approach. Simply  construct the conditional models of $s_1$, from a purist perspective a model can also be constructed of $s_0$, but standard practice here is  usually to take averages over past samples.  Again, there is no need for anything extra or special to `do causal inference'. 

Despite the availability of such a simple method, a significant contextual bandit literature proposes to use a non-Bayesian estimator in order to approximate Equation \ref{valuefunc} based on the Horvitz-Thompson or Inverse Propensity Score estimator also called the inverse propensity score estimator \cite{horvitz1952generalization,bottou2013counterfactual}.
The estimator relies on knowing a logging policy $\pi_0(a_1|S_0)$, and in its simplest form is:

\[
\hat{\mathcal{V}}(f(\cdot)) = \frac{1}{N} \sum_n^N U(s_{1_n}) \frac{\boldsymbol{1}\{ a_{1_n}=f(s_{0_n}) \} }{\pi_0(A_1=a_{1_n} |S_0=s_{0_n})}
\]
The estimator is based on applying the importance sampling formula to samples drawn from: $s_1,a_{1},s_0 \sim P(S_1=s_1|A_1=a_0,S_0=s_0)\pi_0(A_1=a_1|S_0=s_0)P(S_0=s_0)$.
While this estimator is un-biased it is famously very high variance.  


Variance problems with the estimator have lead to a large technical literature of ad-hoc fixes.  For example a significant literature proposes modifications of the ratio $w=\frac{\boldsymbol{1}\{ a_{1_n}=f(s_{0_n}) \} }{\pi_0(A_1=a_{1_n} |S_0=s_{0_n})}$ in order to improve variance properties \cite{bottou2013counterfactual,joachims2018deep,swaminathan2015counterfactual,aouali2023exponential}.  In practice optimizing under a modified estimators favors new polices that are similar to old policy, if the original policy was reasonable then this might plausibly perform better than the un-modified estimator, but only under the assumption you  often already knew the correct action, or when in doubt the old policy gets the benefit of that doubt.

Trying to improve this estimator simply by modifying the ratio $w$ is a  limited perspective, and some methods propose instead to use information about  items \cite{saito2022off,sachdeva2023off} similarities.   These methods make a large step in the complexity of the problem they can handle, but  are cumbersome in construction compared to a Bayesian approach that can cleanly incorporate not only information between actions, but also between contexts and between contexts and actions.

This estimator and the related Robbins-Ritov examples \cite{ritov2014bayesian} is at the center of an \emph{epic debate}  \cite{fanli} between Noble Prize laureate Christopher Sims \cite{sims2006example,simstalk} and renowned statistician Larry Wasserman \cite{robinsrobins,asimple2012robins,robins2000foundations,wassermantalk}.  The (uncomfortable) position advanced here is to disagree with both Sims and Wasserman as both of their approaches use the propensity score $\pi_0(a|S_0)$, which is a flagrant violation of conditionality \cite{berger1988likelihood}.  

Experimentally Horvitz-Thompson based estimators are sometimes proposed with a recommendation use case, but they are only bench marked on extremely artificial problems using the supervised to bandit transform \cite{joachims2018deep}.  This 
results in problems with tiny action sets and enormous treatment effects that are very easy to perform well on, with little relevance to real recommendation tasks.

\section{Shooting in the Dark Method}
\label{shootinginthedark}

Having explained the difficulties in a direct implementation of the Bayesian decision theoretic approach to recommendation, even if strong contextual bandit assumptions are made, let's now consider what practitioners typically do.  They give up, and  \emph{shoot in the dark}.


The shooting in the dark method proceeds by directly optimizing $f_{1:T}(\cdot)$ wrt to a heuristic method of performance.  That is a pseudo-utility function is proposed:

\begin{align}
f^*_{1:T} = {\rm argmax} ~ \tilde{U}(f_{1:T})
\label{darkloss}
\end{align}

Often the utility function will have no access to the actual state, action data $\mathcal{D}=\{ s_{0:T_u}^u,a_{1:T_u}^u\}_{u=1}^M$ and the pseudo-utility may be defined on unrelated data e.g. collaborative filtering or content.  For example a language model may be used to determine if  there is a `similarity' between $a_1$ and $s_0$,  There are two main choices required in order to formulate Equation \ref{darkloss}, the data used and the optimization problem.  Collaborative filtering is one popular source of a proxy reward signal \cite{harper2015movielens}, perhaps the frequency that two elements co-occur can be used to suggest an affinity between $a_1$ and $s_0$. Formulating the pseudo-utility in terms of preferences is a popular optimization framework \cite{rendle2012bpr}.
These choices are consequential, but having discarded theory are guided by intuition and checking that intuition using A/B testing.



\section{Large Scale Click Models}
\label{pclick}

While the shooting in the dark method can be an end-to-end system, often it is used as a candidate generating mechanism, with a final ranker used to select the final recommendation(s) \cite{hron2021component}.  
Typically these models are trained using maximum likelihood, and the `hashing trick'
\cite{chapelle2014simple,weinberger2009feature}.  
In some ways these models resemble reward optimizing models, but limitations of the tooling make this not really the case.

Recommendation is founded upon obtaining accurate regression estimates of the cross between the context and the action.  Unfortunately the likelihood for most of the parameter space of interaction terms is completely flat, making point estimation an extremely unsatisfactory summary of the posterior.

This difficulty is handled using two heuristics: firstly `feature engineering' is used.  Instead of the full context-action cross, the parameter space is restricted, concentrating the reward signal; secondly
the candidate set of actions is often greatly restricted for each context, further concentrating the likelihood.  These ad-hoc fixes do give some validity to claims  that the regression (or direct method)  suffers from strong biases  \cite{beygelzimer2009offset}. However, this issue lies with the ad-hoc fixes not the direct method.

\section{Proposal: Deep Models, but not to go deep}
\label{deepbutnottogodeep}

Standard production ML supervised learning tools are ideal for situations where a) the model $P(S_1=s_1|S_0=s_0,A_1=a_1,\theta_1)$ is restricted so that $s_1$ is a scalar/reward and contains no additional information; and $a_1$ is fixed dimensional (e.g contains single recommendation); and  the likelihood concentrates such that point estimation is an adequate summary of the posterior.  None of these are requirements are true in real systems making standard tooling inadequate and motivating the adoption of heuristics.
Deep learning however allows us to handle  these limitations cleanly.  In short we will use deep learning, \emph{but not to go deep}.

\subsection{Use All The Information}


While reducing recommendation to a contextual bandit is a reasonable approach, restrictions of current tooling dispose of more information than is necessary.  That is an expected reward model of the form:
$E(v|S_0=s_0,A_1=a_1,\theta_1)$ is estimated.  This ignores all non reward information in $v=U(s_1)$.  Standard tooling also fails to correctly model that in most cases $a_1$ is a slate of many recommendations.  With the availability of production ready deep learning tooling it is no longer necessary to ignore this valuable information.


One instance of valuable non-reward information contained within $s_1$ is preference information.  If multiple items were available to select at once, the reward says one of them were clicked, but there is additional preference information, which ones was clicked, which ones were not clicked.  As discussed previously with respect to the BPR loss \cite{rendle2012bpr}, this preference information can be quite informative.  



A widely known model of the form described is the \emph{cascade model} \cite{chuklin2022click} which contains both a preference and reward model of the form $P(S_1=s_1|S_0=s_0,A_1=a_1,\theta_1)$, but it is a bespoke model for search and not applicable for most recommendation settings.
A more recent development that is more appropriate for  recommendation problems is the  Probabilistic Rank and Reward model \cite{aouali2023probabilistic}.  This model  allows $s_1$ to contain not just a reward, but what was clicked (preference information), it also allows $s_0$ to contain the users interests (and also how engaged the user is), and $a_1$ to contain the full slate of recommended items.  This model does make some assumptions, primarily that the recommendations do not interact in complicated ways.  While relaxing this assumption is conceivable it would result in both an explosion in model dimension (and variance) and also make solving the recommendation problem in Equation \ref{banditopt} infeasible.

A very positive consequence of settling on an appropriate likelihood model  is that the likelihood now contains \emph{all the information in the sample/training data}.  Debates about what is the correct loss are now over.

\subsection{Variational Approximations of Bayesian Inference}

Section \ref{pclick} highlighted the limitation of using maximum likelihood to train the reward model of an (extended) contextual bandit. The fundamental difficulty emerges from the likelihood function itself, when the model is parameterized with the complete interaction effect of the context and the action (both of which are very large in dimension); the likelihood will inevitably be flat over vast regions of parameter space.  In other words: the likelihood contains no information at all about action, context combinations you never tried.  
The ad-hoc fixes widely used in industry involve using feature engineering to restrict the parameter space, and limiting the candidates for each context.
These  heuristics are adopted because the standard production tools implement maximum likelihood, rather than approximate Bayesian inference.  Fortunately, Deep Learning tooling again comes to the rescue  by enabling easy implementation of variational Bayes with techniques such as the local re-parameterization trick \cite{kingma2015variational}.

Not only does Bayes allow handling the very uneven likelihood signal, it also allows incorporating strong prior information.  The non-reward signal that the shooting in the dark method uses, can also be used to create embeddings that identify similarities between contexts ($s_0$ and $s_0'$), similarities between actions ($a_1$ and $a_1'$) and similarities between contexts and actions $s_0$ and $a_1$.  This can be formalized by using a matrix normal prior distribution as demonstrated in \cite{sakhi2020blob}.  The strength of this approach is that the likelihood will take precedence of non-reward signals when it is available, but when it is not prior information will step in to make the best possible inference over the space of all possible context, action pairs.


\section{Conclusion}
\label{conclusion}

In this paper  a number of academic approaches to improving recommender systems were surveyed.  It was argued that much academic work on off-policy estimation, causality and attribution is misdirected.
In contrast \emph{the shooting in the dark} method is deserving of respect.  The problems facing a principled approach to reward optimizing recommendation are formidable.  It therefore isn't surprising that a guess and check method based on offline heuristics has dominated practice.  Also the ideas from this area can help us build informative priors, assisting in developing a truly reward optimizing approach.



However, as deep learning tooling become more robust and available this anti-Utopian perspective is becoming harder to justify.  I is  now within our reach to specifically target and optimize for A/B test performance.  The impact of deep learning tools on recommendation should not be to go deep, but to build bespoke models that learn directly from the reward signal and use Bayes to handle the signals unevenness and to augment with weak reward signal with strong prior information.  Both constructing bespoke models, and approximate Bayesian inference were difficult to build and productionize in the not so recent past, thanks to deep learning this is no longer so.   This opens the possibility of building truly reward optimizing recommender systems.

\section{Impact Statement}
Recommender systems are pervasive throughout society and the improvement in machine learning algorithms for recommendation will consequently have society wide impacts.  This paper is neutral on what the goals of these algorithms should be.  It is important that operators of recommender systems engage with society at large when selecting what long term metrics the algorithms optimize.

\bibliographystyle{plainnat}
\bibliography{refs}

\begin{thebibliography}{52}
\providecommand{\natexlab}[1]{#1}
\providecommand{\url}[1]{\texttt{#1}}
\expandafter\ifx\csname urlstyle\endcsname\relax
  \providecommand{\doi}[1]{doi: #1}\else
  \providecommand{\doi}{doi: \begingroup \urlstyle{rm}\Url}\fi

\bibitem[Aouali et~al.(2023{\natexlab{a}})Aouali, Brunel, Rohde, and
  Korba]{aouali2023exponential}
Imad Aouali, Victor-Emmanuel Brunel, David Rohde, and Anna Korba.
\newblock Exponential smoothing for off-policy learning.
\newblock \emph{arXiv preprint arXiv:2305.15877}, 2023{\natexlab{a}}.

\bibitem[Aouali et~al.(2023{\natexlab{b}})Aouali, Hammou, Ivanov, Sakhi, Rohde,
  and Vasile]{aouali2023probabilistic}
Imad Aouali, Achraf Ait~Sidi Hammou, Sergey Ivanov, Otmane Sakhi, David Rohde,
  and Flavian Vasile.
\newblock Probabilistic rank and reward: A scalable model for slate
  recommendation.
\newblock 2023{\natexlab{b}}.

\bibitem[Bellman(1966)]{bellman1966dynamic}
Richard Bellman.
\newblock Dynamic programming.
\newblock \emph{Science}, 153\penalty0 (3731):\penalty0 34--37, 1966.

\bibitem[Berger and Wolpert(1988)]{berger1988likelihood}
James~O Berger and Robert~L Wolpert.
\newblock The likelihood principle.
\newblock IMS, 1988.

\bibitem[Bernardo and Smith(2009)]{bernardo2009bayesian}
Jos{\'e}~M Bernardo and Adrian~FM Smith.
\newblock \emph{Bayesian theory}, volume 405.
\newblock John Wiley \& Sons, 2009.

\bibitem[Betlei et~al.(2023)Betlei, Vladimirova, Sebbar, Urien, Rahier, and
  Heymann]{betlei2023maximizing}
Artem Betlei, Mariia Vladimirova, Mehdi Sebbar, Nicolas Urien, Thibaud Rahier,
  and Benjamin Heymann.
\newblock Maximizing the success probability of policy allocations in online
  systems.
\newblock \emph{arXiv preprint arXiv:2312.16267}, 2023.

\bibitem[Beygelzimer and Langford(2009)]{beygelzimer2009offset}
Alina Beygelzimer and John Langford.
\newblock The offset tree for learning with partial labels.
\newblock In \emph{Proceedings of the 15th ACM SIGKDD international conference
  on Knowledge discovery and data mining}, pages 129--138, 2009.

\bibitem[Bottou et~al.(2013)Bottou, Peters, Qui{\~n}onero-Candela, Charles,
  Chickering, Portugaly, Ray, Simard, and Snelson]{bottou2013counterfactual}
L{\'e}on Bottou, Jonas Peters, Joaquin Qui{\~n}onero-Candela, Denis~X Charles,
  D~Max Chickering, Elon Portugaly, Dipankar Ray, Patrice Simard, and
  Ed~Snelson.
\newblock Counterfactual reasoning and learning systems: The example of
  computational advertising.
\newblock \emph{Journal of Machine Learning Research}, 14\penalty0 (11), 2013.

\bibitem[Chapelle et~al.(2014)Chapelle, Manavoglu, and
  Rosales]{chapelle2014simple}
Olivier Chapelle, Eren Manavoglu, and Romer Rosales.
\newblock Simple and scalable response prediction for display advertising.
\newblock \emph{ACM Transactions on Intelligent Systems and Technology (TIST)},
  5\penalty0 (4):\penalty0 1--34, 2014.

\bibitem[Chen et~al.(2019)Chen, Beutel, Covington, Jain, Belletti, and
  Chi]{chen2019top}
Minmin Chen, Alex Beutel, Paul Covington, Sagar Jain, Francois Belletti, and
  Ed~H Chi.
\newblock Top-k off-policy correction for a reinforce recommender system.
\newblock In \emph{Proceedings of the Twelfth ACM International Conference on
  Web Search and Data Mining}, pages 456--464, 2019.

\bibitem[Chen et~al.(2021)Chen, Yao, McAuley, Zhou, and Wang]{chen2021survey}
Xiaocong Chen, Lina Yao, Julian McAuley, Guanglin Zhou, and Xianzhi Wang.
\newblock A survey of deep reinforcement learning in recommender systems: A
  systematic review and future directions.
\newblock \emph{arXiv preprint arXiv:2109.03540}, 2021.

\bibitem[Chuklin et~al.(2022)Chuklin, Markov, and De~Rijke]{chuklin2022click}
Aleksandr Chuklin, Ilya Markov, and Maarten De~Rijke.
\newblock \emph{Click models for web search}.
\newblock Springer Nature, 2022.

\bibitem[Dawid(2021)]{dawid2021decision}
Philip Dawid.
\newblock Decision-theoretic foundations for statistical causality.
\newblock \emph{Journal of Causal Inference}, 9\penalty0 (1):\penalty0 39--77,
  2021.

\bibitem[De~Finetti(1937)]{de1937prevision}
Bruno De~Finetti.
\newblock La pr{\'e}vision: ses lois logiques, ses sources subjectives.
\newblock In \emph{Annales de l'institut Henri Poincar{\'e}}, volume~7, pages
  1--68, 1937.

\bibitem[Gao et~al.(2022)Gao, Zheng, Wang, Feng, He, and Li]{gao2022causal}
Chen Gao, Yu~Zheng, Wenjie Wang, Fuli Feng, Xiangnan He, and Yong Li.
\newblock Causal inference in recommender systems: A survey and future
  directions.
\newblock \emph{ACM Transactions on Information Systems}, 2022.

\bibitem[Harper and Konstan(2015)]{harper2015movielens}
F~Maxwell Harper and Joseph~A Konstan.
\newblock The movielens datasets: History and context.
\newblock \emph{Acm transactions on interactive intelligent systems (tiis)},
  5\penalty0 (4):\penalty0 1--19, 2015.

\bibitem[Horvitz and Thompson(1952)]{horvitz1952generalization}
Daniel~G Horvitz and Donovan~J Thompson.
\newblock A generalization of sampling without replacement from a finite
  universe.
\newblock \emph{Journal of the American statistical Association}, 47\penalty0
  (260):\penalty0 663--685, 1952.

\bibitem[Hron et~al.(2021)Hron, Krauth, Jordan, and
  Kilbertus]{hron2021component}
Jiri Hron, Karl Krauth, Michael Jordan, and Niki Kilbertus.
\newblock On component interactions in two-stage recommender systems.
\newblock \emph{Advances in neural information processing systems},
  34:\penalty0 2744--2757, 2021.

\bibitem[Ie et~al.(2019)Ie, Jain, Wang, Narvekar, Agarwal, Wu, Cheng, Chandra,
  and Boutilier]{ie2019slateq}
Eugene Ie, Vihan Jain, Jing Wang, Sanmit Narvekar, Ritesh Agarwal, Rui Wu,
  Heng-Tze Cheng, Tushar Chandra, and Craig Boutilier.
\newblock Slateq: A tractable decomposition for reinforcement learning with
  recommendation sets.
\newblock 2019.

\bibitem[Imbens and Rubin(2010)]{imbens2010rubin}
Guido~W Imbens and Donald~B Rubin.
\newblock Rubin causal model.
\newblock In \emph{Microeconometrics}, pages 229--241. Springer, 2010.

\bibitem[Initiative(2022)]{msi}
Marketing~Science Initiative.
\newblock Research priorities.
\newblock
  \href{https://www.msi.org/wp-content/uploads/2022/10/MSI-2022-24-Research-Priorities-Final.pdf}{Report},
  2022.

\bibitem[Joachims et~al.(2018)Joachims, Swaminathan, and
  De~Rijke]{joachims2018deep}
Thorsten Joachims, Adith Swaminathan, and Maarten De~Rijke.
\newblock Deep learning with logged bandit feedback.
\newblock In \emph{International Conference on Learning Representations}, 2018.

\bibitem[Kadane(2020)]{kadane2020principles}
Joseph~B Kadane.
\newblock \emph{Principles of uncertainty}.
\newblock CRC press, 2020.

\bibitem[Kingma et~al.(2015)Kingma, Salimans, and
  Welling]{kingma2015variational}
Durk~P Kingma, Tim Salimans, and Max Welling.
\newblock Variational dropout and the local reparameterization trick.
\newblock \emph{Advances in neural information processing systems}, 28, 2015.

\bibitem[Kohavi et~al.(2009)Kohavi, Longbotham, Sommerfield, and
  Henne]{kohavi2009controlled}
Ron Kohavi, Roger Longbotham, Dan Sommerfield, and Randal~M Henne.
\newblock Controlled experiments on the web: survey and practical guide.
\newblock \emph{Data mining and knowledge discovery}, 18:\penalty0 140--181,
  2009.

\bibitem[Kohavi et~al.(2013)Kohavi, Deng, Frasca, Walker, Xu, and
  Pohlmann]{kohavi2013online}
Ron Kohavi, Alex Deng, Brian Frasca, Toby Walker, Ya~Xu, and Nils Pohlmann.
\newblock Online controlled experiments at large scale.
\newblock In \emph{Proceedings of the 19th ACM SIGKDD international conference
  on Knowledge discovery and data mining}, pages 1168--1176, 2013.

\bibitem[Lad(1996)]{lad1996operational}
Frank Lad.
\newblock \emph{Operational subjective statistical methods: a mathematical,
  philosophical and historical introduction}.
\newblock 1996.

\bibitem[Lattimore and Rohde(2019)]{lattimore2019replacing}
Finnian Lattimore and David Rohde.
\newblock Replacing the do-calculus with bayes rule.
\newblock \emph{arXiv preprint arXiv:1906.07125}, 2019.

\bibitem[Li(2022)]{fanli}
Fan Li.
\newblock Propensity score in {B}ayesian causal inference: why, why not, and
  how?
\newblock \href{https://www.youtube.com/watch?v=_BjkF2nl7dg}{Video}, 2022.

\bibitem[Pearl(1995)]{pearl1995causal}
Judea Pearl.
\newblock Causal diagrams for empirical research.
\newblock \emph{Biometrika}, 82\penalty0 (4):\penalty0 669--688, 1995.

\bibitem[Peierls(1960)]{peierls1960wolfgang}
Rudolf~Ernst Peierls.
\newblock Wolfgang {E}rnst {P}auli, 1900-1958, 1960.

\bibitem[Rendle et~al.(2012)Rendle, Freudenthaler, Gantner, and
  Schmidt-Thieme]{rendle2012bpr}
Steffen Rendle, Christoph Freudenthaler, Zeno Gantner, and Lars Schmidt-Thieme.
\newblock {BPR}: {B}ayesian personalized ranking from implicit feedback.
\newblock \emph{arXiv preprint arXiv:1205.2618}, 2012.

\bibitem[Ritov et~al.(2014)Ritov, Bickel, Gamst, and Kleijn]{ritov2014bayesian}
Ya’acov Ritov, Peter~J Bickel, Anthony~C Gamst, and Bastiaan Jan~Korneel
  Kleijn.
\newblock The {B}ayesian analysis of complex, high-dimensional models: Can it
  be {CODA}?
\newblock 2014.

\bibitem[Robert et~al.(2007)]{robert2007bayesian}
Christian~P Robert et~al.
\newblock \emph{The Bayesian choice: from decision-theoretic foundations to
  computational implementation}, volume~2.
\newblock Springer, 2007.

\bibitem[Robins()]{robinsrobins}
James Robins.
\newblock Robins and {W}asserman respond to a {N}obel prize winner.

\bibitem[Robins and Wasserman(2000)]{robins2000foundations}
James Robins and Larry Wasserman.
\newblock The foundations of statistics: A vignette.
\newblock \emph{J. Amer. Statist. Assoc}, 95:\penalty0 1340--1346, 2000.

\bibitem[Rohde(2022{\natexlab{a}})]{rohde2022causal}
David Rohde.
\newblock Causal inference, is just inference: A beautifully simple idea that
  not everyone accepts.
\newblock In \emph{I (Still) Can't Believe It's Not Better! Workshop at NeurIPS
  2021}, pages 75--79. PMLR, 2022{\natexlab{a}}.

\bibitem[Rohde(2022{\natexlab{b}})]{rohdecaus}
David Rohde.
\newblock Causal inference is inference.
\newblock \href{https://www.youtube.com/watch?v=hoFRVJLnLQc}{Video},
  2022{\natexlab{b}}.

\bibitem[Rubin(1986)]{rubin1986statistics}
Donald~B Rubin.
\newblock Statistics and causal inference: Comment: Which ifs have causal
  answers.
\newblock \emph{Journal of the American Statistical Association}, 81\penalty0
  (396):\penalty0 961--962, 1986.

\bibitem[Sachdeva et~al.(2023)Sachdeva, Wang, Liang, Kallus, and
  McAuley]{sachdeva2023off}
Noveen Sachdeva, Lequn Wang, Dawen Liang, Nathan Kallus, and Julian McAuley.
\newblock Off-policy evaluation for large action spaces via policy convolution.
\newblock \emph{arXiv preprint arXiv:2310.15433}, 2023.

\bibitem[Saito and Joachims(2022)]{saito2022off}
Yuta Saito and Thorsten Joachims.
\newblock Off-policy evaluation for large action spaces via embeddings.
\newblock \emph{arXiv preprint arXiv:2202.06317}, 2022.

\bibitem[Sakhi et~al.(2020)Sakhi, Bonner, Rohde, and Vasile]{sakhi2020blob}
Otmane Sakhi, Stephen Bonner, David Rohde, and Flavian Vasile.
\newblock {BLOB}: A probabilistic model for recommendation that combines
  organic and bandit signals.
\newblock In \emph{Proceedings of the 26th ACM SIGKDD International Conference
  on Knowledge Discovery \& Data Mining}, pages 783--793, 2020.

\bibitem[Sims(2006)]{sims2006example}
Christopher Sims.
\newblock On an example of larry wasserman.
\newblock \emph{online manuscript, available from http://sims. princeton.
  edu/yftp/WassermanExmpl/WassermanComment. pdf}, 2\penalty0 (10), 2006.

\bibitem[Sims(2012)]{asimple2012robins}
Christopher Sims.
\newblock {R}obins-{W}asserman, round n.
\newblock 2012.

\bibitem[Sims(2022)]{simstalk}
Christopher Sims.
\newblock Christopher sims - large parameter spaces and weighted data: A
  {B}ayesian perspective.
\newblock \href{https://www.youtube.com/watch?v=3JMwAtONsws}{Video}, 2022.

\bibitem[Singal et~al.(2019)Singal, Besbes, Desir, Goyal, and
  Iyengar]{singal2019shapley}
Raghav Singal, Omar Besbes, Antoine Desir, Vineet Goyal, and Garud Iyengar.
\newblock Shapley meets uniform: An axiomatic framework for attribution in
  online advertising.
\newblock In \emph{The World Wide Web Conference}, pages 1713--1723, 2019.

\bibitem[Sutton and Barto(2018)]{sutton2018reinforcement}
Richard~S Sutton and Andrew~G Barto.
\newblock \emph{Reinforcement learning: An introduction}.
\newblock MIT press, 2018.

\bibitem[Swaminathan and Joachims(2015)]{swaminathan2015counterfactual}
Adith Swaminathan and Thorsten Joachims.
\newblock Counterfactual risk minimization: Learning from logged bandit
  feedback.
\newblock In \emph{International Conference on Machine Learning}, pages
  814--823. PMLR, 2015.

\bibitem[Wasserman(2022)]{wassermantalk}
Larry Wasserman.
\newblock Problems with {B}ayesian causal inference.
\newblock \href{https://www.youtube.com/watch?v=sZyyaNdvfto}{Video}, 2022.

\bibitem[Watkins and Dayan(1992)]{watkins1992q}
Christopher~JCH Watkins and Peter Dayan.
\newblock Q-learning.
\newblock \emph{Machine learning}, 8:\penalty0 279--292, 1992.

\bibitem[Weinberger et~al.(2009)Weinberger, Dasgupta, Langford, Smola, and
  Attenberg]{weinberger2009feature}
Kilian Weinberger, Anirban Dasgupta, John Langford, Alex Smola, and Josh
  Attenberg.
\newblock Feature hashing for large scale multitask learning.
\newblock In \emph{Proceedings of the 26th annual international conference on
  machine learning}, pages 1113--1120, 2009.

\bibitem[Williams(1992)]{williams1992simple}
Ronald~J Williams.
\newblock Simple statistical gradient-following algorithms for connectionist
  reinforcement learning.
\newblock \emph{Machine learning}, 8:\penalty0 229--256, 1992.

\end{thebibliography}
\end{document}